\documentclass[12pt]{article}
\usepackage{amssymb,amsmath,epsfig}

\begin{document}

\title{\bf Gravitational Charged Perfect Fluid Collapse in Friedmann Universe Models}

\author{M. Sharif \thanks{msharif@math.pu.edu.pk} and G. Abbas
\thanks{abbasg91@yahoo.com}\\
Department of Mathematics, University of the Punjab,\\
Quaid-e-Azam Campus, Lahore-54590, Pakistan.}

\date{}
\maketitle

\begin{abstract}
This paper is devoted to study the gravitational charged perfect
fluid collapse in the Friedmann universe models with cosmological
constant. For this purpose, we assume that the electromagnetic
field is so weak that it does not introduce any distortion into
the geometry of the spacetime. The results obtained from the
junction conditions between the Friedmann and the
Reissner-Nordstr$\ddot{o}$m de-Sitter spacetimes are used to solve
the field equations. Further, the singularity structure and mass
effects of the collapsing system on time difference between the
formation of apparent horizons and singularity have been studied.
This analysis provides the validity of Cosmic Censorship
Hypothesis. It is found that the electric field affects the area
of apparent horizons and their time of formation.
\end{abstract}


\section{Introduction}

Gravitational collapse of a massive star is the result of its self
gravity. It occurs when the internal nuclear fuel of the star
fails to supply sufficiently high pressure to counter-balance
gravity. Gravitational collapse is one of the most important
problems in general relativity. According to the singularity
theorems \cite{1} there exist spacetime singularities in generic
gravitational collapse. It has been an interesting problem to
determine the nature of spacetime singularity. The cosmic
censorship hypothesis (CCH) \cite{2} says that singularities
appearing in gravitational collapse are always clothed by the
event horizon.

The final fate of gravitational collapse of the massive star
depends upon the choice of initial data and equation of state.
Many efforts have been made to check its credibility but no final
conclusion is drawn. For this purpose, Virbhadra et al. \cite{3}
introduced a new theoretical tool using the gravitational lensing
phenomena. Also, Virbhadra and Ellis \cite{4} studied the
Schwarzschild black hole lensing and found that the relativistic
images would confirm the Schwarzschild geometry close to the event
horizon. The same authors \cite{5} analyzed the gravitational
lensing by a naked singularity and classified it as weak naked
singularity and strong naked singularity. In a recent paper
\cite{6}, Virbhadra used the gravitational lensing phenomena to
find the improved form of the CCH.

Oppenheimer and Snyder \cite{7} studied dust collapse for the
first time and showed that singularity is neither locally or
globally naked. This means that they found black hole as a final
fate of the dust collapse. Eardely and Smarr \cite{8} found that
inhomogeneous model undergoes to gravitational collapse by forming
a singularity that can be either locally or globally naked.

There has been a growing interest to study gravitational collapse
in the presence of perfect fluid and other general physical form
of the fluid. Misner and Sharp \cite{9} extended the pioneer work
for the perfect fluid. Vaidya \cite{10} and Santos \cite{11} used
the idea of outgoing radiation of the collapsing body and also
included the dissipation in the source by allowing the radial heat
flow. Markovic and Shapiro \cite{12} generalized the pioneer work
with positive cosmological constant. Lake \cite{13} extended it
for both positive and negative cosmological constant. Sharif and
Ahmad \cite{14}-\cite{17} extended spherically symmetric
gravitational collapse with positive cosmological constant for
perfect fluid. The same authors \cite{18} have also investigated
plane symmetric gravitational collapse using junction conditions
which has been extended to spherically symmetric gravitational
collapse \cite{19}.

The behavior of electromagnetic field in strong gravitational
field has been the subject of interest for the researchers over
the past decades. According to Thirukkanesh and Maharaj \cite{20},
the inclusion of electromagnetic field in gravitational collapse
predicts that the gravitational attraction is counterbalanced by
the Coulomb repulsive force along with the pressure gradient.
Sharma et al. \cite{21} have concluded that electromagnetic field
affects the value of red-shift, luminosity and mass of the
relativistic compact objects. Nath et al. \cite{22} have studied
the gravitational collapse of non-viscous, heat conducting fluid
in the presence of electromagnetic field. They concluded that
electromagnetic field reduces pressure and favors the formation of
naked singularity.

Recently, we have studied the effect of electromagnetic field on
the gravitational collapse by taking spherically symmetric
spacetime as interior region and Reissner-Nordstr$\ddot{o}$m as
exterior region of the star \cite{23}. The present article
investigates the previous work by taking the Friedmann universe
models in the interior of star. In order to preserve the generic
properties of the Friedmann universe models in the presence of
electromagnetic field, we follow \cite{24,25} and assume that
electromagnetic field is weak relative to matter, i.e., if $E^2$
is the electromagnetic field contribution in the system then
$E^2<<\rho$. The main objectives of this work are the following:
\begin{itemize}
\item To study the physical interpretation of electromagnetic
field and cosmological constant on gravitational collapse in the
Friedmann universe models.
\item  To see the validity of CCH in this framework.
\end{itemize}

The plan of the paper is as follows: In the next section, the
junction conditions are given. We discuss the solution of the
Einstein-Maxwell field equations in section \textbf{3}. The
apparent horizons and their physical significance are presented in
section \textbf{4}. Section \textbf{5} presents the singularity
analysis. We conclude our discussion in the last section.

The geometrized units (i.e., the gravitational constant $G$=1 and
speed of light in vacuum $c=1$ so that $M\equiv\frac{MG}{c^2}$ and
$\kappa\equiv\frac{8\pi G}{c^4}=8\pi$) are used. All the Latin and
Greek indices vary from 0 to 3, otherwise, it will be mentioned.

\section{Junction Conditions}

We derive conditions for the smooth matching of two regions
(interior and exterior of a star) on the surface of discontinuity.
For this purpose, we assume that $\Sigma$ be a timelike $3D$
hypersurface which divides two $4D$ manifolds $V^-$ and $V^+$
respectively. The interior manifold  is taken as the Friedmann
model
\begin{equation}\label{1}
ds_-^2=dt^2-a(t)^2[d\chi^2-f^2(d\theta^2+\sin\theta^2d\phi^2)],
\end{equation}
where $f_k (\chi)$ is defined as
\begin{equation}
f(\chi)=\begin{cases}
\sin\chi, &k=1\\
\chi,     &k=0,\\
\sinh\chi,  &k=-1,
\end{cases}
\end{equation}
$k=1,0,-1$ correspond to closed, flat and open models
respectively. $\chi (0\leq\chi<\infty$ for open and closed but
$0\leq\chi<\pi$ for flat) is the hyper-spherical angle and $a(t)$
is the scale factor. Further, $\chi$ is related to radial
coordinate $r$ as follows: $r=\sin\chi$ (closed),  $r=\chi$ (flat)
and $r=\sinh \chi$ (open). The Reissner-Nordstr$\ddot{o}$m
de-Sitter spacetime is taken as the exterior manifold
\begin{equation}\label{2}
ds_+^2=ZdT^2-\frac{1}{Z}dR^2-R^2(d\theta^2+\sin\theta^2d\phi^2),
\end{equation}
where
\begin{equation}\label{3}
Z(R)=1-\frac{2M}{R}+\frac{Q^2}{R^2}-\frac{\Lambda}{3}R^2,
\end{equation}
$M$ and $\Lambda$ are constants and $Q$ is the charge.

The junction conditions are given as follows \cite{26}:
\begin{enumerate}
\item The continuity of first fundamental form over $\Sigma$ gives
\begin{equation}\label{4}
(ds^2_-)_{\Sigma}=(ds^2_+)_{\Sigma}=ds^2_{\Sigma}.
\end{equation}
\item The continuity of second fundamental form (extrinsic curvature)
over $\Sigma$ yields
\begin{equation}\label{5}
[K_{ij}]=K^+_{ij}-K^-_{ij}=0, \quad(i,j=0,2,3)
\end{equation}
where $K_{ij}$ is the extrinsic curvature defined as
\end{enumerate}
\begin{equation}\label{6}
K^{\pm}_{ij}=-n^{\pm}_{\sigma}(\frac{{\partial}^2x^{\sigma}_{\pm}}
{{\partial}{\xi}^i{\partial}{\xi}^j}+{\Gamma}^{\sigma}_{{\mu}{\nu}}
\frac{{{\partial}x^{\mu}_{\pm}}{{\partial}x^{\nu}_{\pm}}}
{{\partial}{\xi}^i{\partial}{\xi}^j}),\quad({\sigma},
{\mu},{\nu}=0,1,2,3).
\end{equation}
Here $\xi^0= t$, $\xi^2=\theta$, $\xi^3= \phi$ are the
corresponding parameters on ${\Sigma }$, $x^{\sigma}_{\pm}$  stand
for coordinates in $V^{\pm}$,  the Christoffel symbols
$\Gamma^{\sigma}_{{\mu}{\nu}}$ are calculated from the interior or
exterior spacetimes and $n^{\pm}_{\sigma}$ are the components of
outward unit normals to ${\Sigma}$ in the coordinates
$x^{\sigma}_{\pm}$.

The equation of hypersurface in terms of interior spacetime $V^-$
coordinates is
\begin{equation}\label{8}
f_-(\chi,t)=\chi-\chi_{\Sigma}=0,
\end{equation}
where $\chi_{\Sigma}$ is a constant as $\Sigma$ is a comoving
surface forming the boundary of interior matter. Also, the
equation of hypersurface in terms of exterior spacetime $V^+$
coordinates is given by
\begin{equation}\label{9}
f_+(R,T)=R-R_{\Sigma}(T)=0.
\end{equation}
When we make use of Eq.(\ref{8}) in Eq.(\ref{1}), the metric on
$\Sigma$ takes the form
\begin{equation}\label{10}
(ds_-^2)_\Sigma={dt^2-a(t)^2f(\chi_\Sigma)(d\theta^2+\sin\theta^2d\phi^2)}.
\end{equation}
Also, Eqs.(\ref{9}) and (\ref{2}) yield
\begin{equation}\label{11}
(ds_+^2)_\Sigma=[Z(R_\Sigma)-\frac{1}{Z(R_\Sigma)}
(\frac{dR_\Sigma}{dT})^2]dT^2-R_\Sigma^2(d\theta^2+\sin\theta^2d\phi^2),
\end{equation}
where we assume that
\begin{equation}\label{12}
Z(R_\Sigma)-\frac{1}{Z(R_\Sigma)} (\frac{dR_\Sigma}{dT})^2>0
\end{equation}
so that T is a timelike coordinate. From Eqs.(\ref{4}), (\ref{10})
and (\ref{11}), it follows that
\begin{eqnarray}\label{13}
R_\Sigma=(af)_\Sigma,\\\label{14}
[Z(R_\Sigma)-\frac{1}{Z(R_\Sigma)}
(\frac{dR_\Sigma}{dT})^2]^{\frac{1}{2}}dT=dt .
\end{eqnarray}
Also, from Eqs.(\ref{8}) and (\ref{9}), the outward unit normals
in $V^-$ and $V^+$, respectively, are given by
\begin{eqnarray}\label{15}
n^-_\mu&=&(0,a(t),0,0),\\
\label{16} n^+_\mu&=&(-\dot{R}_\Sigma,\dot{T}, 0,0).
\end{eqnarray}

The components of extrinsic curvature $K^\pm_{ij}$ become
\begin{eqnarray}\label{17}
K^-_{00}&=&0,\\
\label{18}
K^-_{22}&=&\csc^2{\theta}K^-_{33}=({ff'}{a})_\Sigma,\\
\label{19}
K^+_{00}&=&(\dot{R}\ddot{T}-\dot{T}\ddot{R}-\frac{Z}{2}\frac{dZ}{dR}\dot{T}^3
+\frac{3}{2Z}\frac{dZ}{dR}\dot{T}\dot{R}^2)_\Sigma,\\
\label{20} K^+_{22}&=&\csc^2{\theta}
K^+_{33}=(ZR\dot{T})_{\Sigma},
\end{eqnarray}
where dot and prime mean differentiation with respect to $t$ and
$\chi$ respectively. From Eq.(\ref{5}), the continuity of
extrinsic curvature gives
\begin{eqnarray}\label{21}
K^+_{00}&=&0,\\
\label{22} K^+_{22}&=&K^-_{22}.
\end{eqnarray}
Using Eqs.(\ref{17})-(\ref{22}) along with Eqs.(\ref{3}),
(\ref{13}) and (\ref{14}), the junction conditions become
\begin{eqnarray}\label{23}
\dot{(f')}_\Sigma=0,\\
\label{24} M=(\frac{af}{2}-\frac{\Lambda}{6}(af)^3+\frac{Q^2}{2af}
+\frac{{a\dot{a}^2}}{2}{f}^3-\frac{a}{2}{f}{f'}^2)_{\Sigma}.
\end{eqnarray}
Equations (\ref{13}), (\ref{14}), (\ref{23}) and (\ref{24})
provide the necessary and sufficient conditions for the smooth
matching of the two regions over $\Sigma$.

\section{Solution of the Einstein Field Equations}

In this section, we solve the Einstein field equations with
cosmological constant for the Friedmann models containing the
charged perfect fluid as the source of gravitation. The Einstein
field equations with cosmological constant are given by
${\setcounter{equation}{0}}$
\begin{equation}\label{25}
G_{\mu\nu}-{\Lambda}g_{\mu\nu}=\kappa(T_{\mu\nu}+T^{({em})}_{\mu\nu}).
\end{equation}
The energy-momentum tensor for perfect fluid is
\begin{equation}\label{26}
{T_{{\mu}{\nu}}={({\rho}+p)}u_{\mu}u_{\nu}-pg_{\mu\nu}},
\end{equation}
where $\rho$ is the energy density, $p$ is the pressure and
$u_\mu=\delta^0_\mu$ is the four-vector velocity in co-moving
coordinates. $T^{({em})}_{\mu\nu}$ is the energy-momentum tensor
for the electromagnetic field given by
\begin{equation}\label{27}
T^{(em)}_{{\mu}{\nu}}=\frac{1}{4{\pi}}(-g^{{\delta}{\omega}}
F_{{\mu}{\delta}}F_{{\nu}{\omega}}+\frac{1}{4}g_{{\mu}{\nu}}
F_{{\delta}{\omega}}F^{{\delta}{\omega}}).
\end{equation}
With the help of Eqs.(\ref{26}) and (\ref{27}), Eq.(\ref{25})
takes the form
\begin{equation}\label{28}
R_{{\mu}{\nu}}=8\pi[({\rho}+p)u_{\mu}u_{\nu}
+\frac{1}{2}(p-{\rho})g_{{\mu}{\nu}}
+T^{({em})}_{{\mu}{\nu}}-\frac{1}{2}g_{{\mu}{\nu}}T^{({em})}]
-{\Lambda}g_{{\mu}{\nu}}.
\end{equation}

Now we solve the Maxwell's field equations
\begin{eqnarray}\label{29}
F_{\mu\nu}&=&\phi_{\nu,\mu}-\phi_{\mu,\nu},\\\label{30}
F^{\mu\nu}_{}{;\nu}&=&4{\pi}J^{\mu},
\end{eqnarray}
where $\phi_{\mu}$ is the four potential and $J^{\mu}$ is the four
current. Since the charge is at rest in this system, the magnetic
field will be zero. Thus we can choose the four potential and four
current as follows
\begin{eqnarray}\label{31}
\phi_{\mu}=({\phi}(t,r),0,0,0),\quad J^{\mu}={\sigma}u^{\mu},
\end{eqnarray}
where $\sigma$ is charge density. Using Eqs.(\ref{29}) and
(\ref{31}), the non-zero components of the field tensor are given
as follows:
\begin{equation}\label{33}
F_{01}=-F_{10}=-\frac{\partial\phi}{\partial {\chi}}.
\end{equation}
Also, from Eqs.(\ref{30}), (\ref{31}) and (\ref{33}), we have
\begin{eqnarray}\label{34}
\frac{\partial^2\phi}{\partial {\chi}^2}+2
\frac{f'}{f}=4{\pi}{\sigma} a^2,\\ \label{35}
a\frac{\partial^2\phi}{\partial {\chi
\partial{t}}}+{\dot{a}}\frac{\partial\phi}{\partial {\chi}} =0.
\end{eqnarray}

Integration of Eq.(\ref{36}) implies that
\begin{equation}\label{36}
\frac{\partial{\phi}}{\partial {\chi}}= \frac{1}{af^2} q(\chi),
\end{equation}
where
 $q(\chi) = 4{\pi} \int^{\chi}_0\sigma{a^3f^2d{\chi}}$,
is the total charge distribution in the interior spacetime. This
amount of charge is the consequence of law of conservation of
charge, i.e., $J^\mu_{; \mu}=0$. It is clear that Eq.(\ref{35}) is
identically satisfied by Eq.(\ref{36}). The electromagnetic field
intensity is given by
\begin{equation}\label{37}
E=\frac{q}{(af)^2}.
\end{equation}
Equations (\ref{36}) and (\ref{37}) yield
\begin{equation}\label{38}
\frac{\partial{\phi}}{\partial {\chi}}= aE.
\end{equation}
Using Eqs. (\ref{33}) and (\ref{38}), we get

\begin{equation}\label{39}
F_{01}=-F_{10}=-aE.
\end{equation}
The non-zero components of $T^{(em)}_{{\mu}{\nu}}$ and its trace
free form turn out to be
\begin{eqnarray*}
T^{(em)}_{{0}{0}}&=&\frac{1}{8{\pi}}E^2 ,\quad
T^{(em)}_{{1}{1}}=-\frac{1}{8{\pi}}E^2 a^2 ,\quad
T^{(em)}_{{2}{2}}=\frac{1}{8{\pi}}E^2(af)^2,\\
T^{(em)}_{{3}{3}}&=&T^{(em)}_{{2}{2}}\sin^2\theta,\quad
T^{(em)}=0.
\end{eqnarray*}
When we use these values, the field equations (\ref{28}) for the
interior spacetime takes the form
\begin{eqnarray}\label{42}
R_{00}&=&-3\frac{\ddot{a}}{a}=4\pi(\rho+3p)
+E^2-{\Lambda},\\
\label{43} R_{11}&=&-\frac{\ddot{a}}{a}-2 \frac{\dot{a}^2}{a^2}
+\frac{2}{{a}^2}\frac{f''}{f}={4\pi}(p-\rho)+E^2-{\Lambda} ,\\
\label{44} R_{22}&=&-\frac{\ddot{a}}{a}-(\frac{\dot{a}}{a})^2
+\frac{1}{a^2}[\frac{f''}{f}+(\frac{f'}{f})^2- \frac{1}{f^2}]
={4\pi}(p-\rho)-E^2-{\Lambda} ,\\
\label{45} R_{33}&=&{\sin}^2{\theta}R_{22},
\end{eqnarray}

We would like to mention here that all the results are valid for
$E^2 <<\rho$ and hence for stiff matter $(\rho=p),~E^2<<p$.
Integrating Eq.(\ref{23}) with respect to $t$, it follows that
\begin{equation}\label{46}
f'=W,
\end{equation}
where $W=W(\chi)$ is an arbitrary function of $\chi$. The energy
conservation equation
\begin{equation}\label{47}
T^{\nu}_{{\mu};{\nu}}=0
\end{equation}
for the perfect fluid with the interior metric shows that pressure
is a function of $t$ only, i.e.,
\begin{equation}\label{48}
p=p(t).
\end{equation}
Using the values of $f'$ and $p$ from Eqs.(\ref{46}) and
(\ref{48}) in Eqs.(\ref{42})-(\ref{44}), it follows that
\begin{equation}\label{49}
2\frac{\ddot{a}}{a}+(\frac{\dot{a}}{a})^2+\frac{(1-W^2)}{(af)^2}
=\Lambda+{E^2}-8\pi p(t).
\end{equation}
We consider $p$ as a polynomial in $t$ as given by \cite{15}
\begin{equation}\label{50}
p(t)=p_c(\frac{t}{T})^{-s},
\end{equation}
where $T$ is the constant time introduced in the problem due to
physical reason by re-scaling of $t$, $p_c$ and $s$ are positive
constants. Further, for simplicity, we take $s=0$ so that
\begin{equation}\label{51}
p(t)=p_c.
\end{equation}
Now Eq.(\ref{49}) gives
\begin{equation}\label{52}
2\frac{\ddot{a}}{a}+(\frac{\dot{a}}{a})^2+\frac{(1-W^2)}{(af)^2}
=\Lambda+{E^2}-{8\pi}p_c.
\end{equation}
For the static charges $E$ is taken as time independent \cite{31},
so integration of above equation with respect to $t$, yields
\begin{equation}\label{53}
{\dot{a}}^2=\frac{W^2-1}{f^2}+(\Lambda+{E^2}-{8\pi}p_c)\frac{a^2}{3}+2\frac{m}{af^3},
\end{equation}
where $m=m(\chi)$ is an arbitrary function of $\chi$ and is
related to the mass of the collapsing system. Substituting
Eqs.(\ref{46}), (\ref{53}) into Eq.(\ref{42}), we get
\begin{equation}\label{54}
m'=\frac{2E'E}{3}(af)^3+{{a}^3{f'}{f^2}}[4\pi(p_c+{\rho})+2{E^2}].
\end{equation}

For physical reasons, we assume that $(p_c+{\rho})\geqslant0$.
Integrating Eq.(\ref{54}) with respect to $\chi$, we obtain
\begin{equation}\label{55}
m(\chi)=4\pi{a^3}\int^{\chi}_0({\rho}+{p_c}){f'}{f^2}d{\chi}+2\int^{\chi}_0
E^2{f'}{f^2}d{\chi}+\frac{2}{3}a^3\int^{\chi}_0{E'E}f^3d
{\chi}+m_0,
\end{equation}
where $m_0$ is taken equal to zero because of finite distribution
of matter at the origin. The function $m(\chi)$ must be positive
because $m(\chi)<0$ implies negative mass which is not physical.
Using Eqs.(\ref{46}) and (\ref{53}) into the junction condition
Eq.(\ref{24}), it follows that
\begin{equation}\label{56}
M=\frac{Q^2}{2af}+m+\frac{1}{6}(\Lambda+E^2-{8\pi}p_c)(af)^3.
\end{equation}
The total energy $\tilde{M}(\chi,t)$ at time $t$ inside the
hypersurface $\Sigma$ can be evaluated by using the definition of
mass function with the contribution of electromagnetic field for the
Friedmann model, which is given by
\begin{equation}\label{58}
\tilde{M}(\chi,t)=\frac{1}{2}(af)(1+({\dot{a}f})^2-{f'}^2) +
\frac{q^2}{2af}.
\end{equation}
Replacing Eqs.(\ref{46}) and (\ref{53}) in Eq.(\ref{58}), we
obtain
\begin{equation}\label{59}
\tilde{M}(r,t)=m(r)+(\Lambda+{E^2}-{8\pi}p_c)\frac{(af)^3}{6}+
\frac{q^2}{2af}.
\end{equation}
From Eqs.(\ref{56}) and (\ref{59}), it can be found that
$\tilde{M}(r,t)=^{\Sigma}M$ if and only if $q=Q$. This result
provides the necessary and sufficient conditions for the continuity
of mass in the interior and exterior regions over boundary surface
$\Sigma$.

Now we take $(\Lambda+{E^2}-{8\pi}p_c)>0$ such that $E^2<<{8\pi}p$
and assume that
\begin{equation}\label{60}
W(\chi)=1.
\end{equation}
In order to obtain the analytic solutions in closed form, we use
Eqs.(\ref{46}), (\ref{53}) and (\ref{60}) so that
\begin{equation}\label{61}
(af)=(\frac{6m}{\Lambda+{E^2}-{8\pi}p_c})^\frac{1}{3}{\sinh^\frac{2}{3}\alpha(\chi,t)}
\end{equation}
 where
\begin{equation}\label{62}
\alpha(\chi,t)=\frac{\sqrt{3(\Lambda+{E^2}-{8\pi}p_c)}}{2}[t_s(\chi)-t)].
\end{equation}
Here $t_s(\chi)$ is an arbitrary function of $\chi$ and is related
to the time of formation of singularity.

\section{Apparent Horizons}

In this section, we discuss the formation of apparent horizons.
The boundary of two trapped spheres whose outward normals are null
is used to find the apparent horizons. Moreover, we discuss the
the physical significance of apparent horizons i.e., area of
apparent horizons, time difference between apparent horizons and
singularity etc. For the interior spacetime, we find the boundary
of two trapped spheres whose outward normals are null as follows:
${\setcounter{equation}{0}}$
\begin{equation}\label{63}
g^{\mu\nu}(af)_{,\mu} (af)_{,\nu}=\dot{({af})}^2-({f'})^2=0.
\end{equation}
Using Eqs.(\ref{46}) and (\ref{53}) in this equation, we get
\begin{equation}\label{64}
(\Lambda+{E^2}-{8\pi}p_c)(af)^3-3(af)+6m=0.
\end{equation}
When $\Lambda=8\pi p_c-{E^2}$, we have $(af)=2m$. This is called
Schwarzschild horizon. For $m=p_c=K=0$, we have
$(af)=\sqrt{\frac{3}{\Lambda}}$, which is called de-Sitter
horizon. Equation (\ref{64}) can have the following positive roots.\\\\
\textbf{Case (i)}: For
$3m<\frac{1}{\sqrt{(\Lambda+{E^2}-{8\pi}p_c)}}$, we obtain two
horizons
\begin{eqnarray}\label{65}
(af)_c&=&\frac{2}{\sqrt{(\Lambda+{E^2}-{8\pi}p_c)}}\cos\frac{\varphi}{3},\\
\label{66} (af)_b&=&\frac{-1}{\sqrt{(\Lambda+{8\pi}{E^2}-p_c)}}
(\cos\frac{\varphi}{3}-\sqrt{3}\sin\frac{\varphi}{3}),
\end{eqnarray}
where
\begin{equation}\label{67}
\cos\varphi=-3m{\sqrt{(\Lambda+{E^2}-{8\pi}p_c)}}.
\end{equation}
If we take $m=0$, it follows from Eqs.(\ref{65}) and (\ref{66})
that $(af)_c=\sqrt{\frac{3}{(\Lambda+{E^2}-{8\pi}p_c)}}$ and
$(af)_b=0$. $(af)_c$ and $(af)_b$ are called cosmological horizon
and black hole horizons respectively. For $m\neq0$ and
$\Lambda\neq{8\pi}p_c-{E^2}$, $(af)_c$
and $(af)_b$ can be generalized \cite{27} respectively.\\\\
\textbf{Case (ii):} For
$3m=\frac{1}{\sqrt{(\Lambda+{E^2}-{8\pi}p_c)}}$, there is only one
positive root which corresponds to a single horizon i.e.,
\begin{equation}\label{68}
(af)_c=(af)_b=\frac{1}{\sqrt{(\Lambda+{E^2}-{8\pi}p_c)}}=(af)_{cb}.
\end{equation}
This shows that both horizons coincide. The range for the
cosmological and black hole horizon can be written as follows
\begin{equation}\label{69}
0\leq (af)_{b} \leq \frac{1}{\sqrt{(\Lambda+{E^2}-{8\pi}p_c)}}
\leq (af)_{c} \leq \sqrt{\frac{3}{(\Lambda+{E^2}-{8\pi}p_c)}} .
\end{equation}
The black hole horizon has its largest proper area
${4\pi}(af)^2=\frac{4\pi}{(\Lambda+{E^2}-{8\pi}p_c)}$ and
cosmological horizon has its area between
$\frac{4\pi}{(\Lambda+{E^2}-{8\pi}p_c)}$ and
$\frac{12\pi}{(\Lambda+{E^2}-{8\pi}p_c)}$.\\\\
\textbf{Case (iii):} For
$3m>\frac{1}{\sqrt{(\Lambda+{E^2}-{8\pi}p_c)}}$, there are no
positive roots and consequently there are no apparent horizons.

We now calculate the time of formation of the apparent horizon
using Eqs.(\ref{61}) and (\ref{64})
\begin{equation}\label{70}
t_n=t_s-\frac{2}{\sqrt{3(\Lambda+{E^2}-{8\pi}p_c})}\sinh^{-1}
(\frac{(af)_n}{2m}-1)^{\frac{1}{2}}, \quad(n=1,2).
\end{equation}
This implies that
\begin{equation}\label{71}
\frac{(af)_n}{2m}=\cosh^{2}\alpha_n,
\end{equation}
where
$\alpha_n(r,\chi)=\frac{\sqrt{3(\Lambda+{E^2}-{8\pi}p_c)}}{2}[t_s(\chi)-t_n)]$.
Equations (\ref{61}) and (\ref{68}) give $(af)_{c}\geq (af)_{b}$
and $t_{b} \geq t_{c}$ respectively. The inequality $t_{b} \geq
t_{c}$ indicates that the cosmological horizon forms earlier than
the black hole horizon. This condition confirms the formation of
black hole.

The time difference between the formation of cosmological horizon
and singularity and the formation of black hole horizon and
singularity can be found as follows. Using
Eqs.(\ref{65})-(\ref{67}), it follows that
\begin{eqnarray}\label{72}
\frac{d(\frac{(af)_c}{2m})}{dm}&=&\frac{1}{m}(-\frac{\sin\frac{\varphi}{3}}{\sin\varphi}
+\frac{3\cos\frac{\varphi}{3}}{\cos\varphi})<0,\\
\label{73}\frac{d(\frac{(af)_b}{2m})}{dm}
&=&\frac{1}{m}(-\frac{\sin\frac{(\varphi+4\pi)}{3}}{\sin\varphi}
+\frac{3\cos\frac{(\varphi+4\pi)}{3}}{\cos\varphi})>0.
\end{eqnarray}
The time difference between the formation of singularity and
apparent horizons is
\begin{equation}\label{74}
\tau_n=t_s-t_n.
\end{equation}
It follows from Eq.(\ref{71}) that
\begin{equation}\label{75}
\frac{d\tau_n}{d(\frac{Y_n}{2m})}
=\frac{1}{\sinh\alpha_n\cosh\alpha_n{\sqrt{3(\Lambda+{E^2}-{8\pi}p_c)}}}.
\end{equation}
Using Eqs.(\ref{72}) and (\ref{75}), we get
\begin{eqnarray}\label{76}
\frac{d\tau_1}{dm}=\frac{d\tau_1}{d(\frac{(af)_c}{2m})}\frac{d(\frac{(af)_c}{2m})}{dm}
=\frac{1}{m{\sqrt{3(\Lambda+{E^2}-{8\pi}p_c)}}\sinh\alpha_1\cosh\alpha_1}\nonumber\\
\times(-\frac{\sin\frac{\varphi}{3}}{\sin\varphi}
+\frac{3\cos\frac{\varphi}{3}}{\cos\varphi})<0 .
\end{eqnarray}
This means that time interval between the formation of
cosmological horizon and singularity is decreased with the
increase of mass. Similarly, from Eqs.(\ref{73}) and(\ref{75}), we
get
\begin{eqnarray}\label{77}
\frac{d\tau_2}{dm}= \frac{1}{m{\sqrt{3
(\Lambda+{E^2}-{8\pi}p_c)}}\sinh\alpha_2\cosh\alpha_2}\nonumber\\
\times(-\frac{\sin\frac{(\varphi+4\pi)}{3}}{\sin\varphi}
+\frac{3\cos\frac{(\varphi+4\pi)}{3}}{\cos\varphi})>0.
\end{eqnarray}
This indicates that time difference between the formation of black
hole horizon and singularity is increased with the increase of
mass.

\section{Singularity Analysis}

The Riemann tensor is used to determine whether a singularity is
essential or removable. If the curvature becomes infinite at
certain point, then the singularity will be essential otherwise
removable. Many scalars can be constructed from the Riemann tensor
but symmetry assumption can be used to find only a finite number
of independent scalars. Some of these are
$$ R_1=R=g^{ab}R_{ab},\quad R_2=R_{ab}R^{ab},\quad
R_3=R_{abcd}R^{abcd},\quad R_4=R^{ab}_{cd}R_{ab}^{cd}.$$ Here, we
give the analysis for the first invariant commonly known as the
Ricci scalar. For the Friedmann  model, it is given as

\begin{equation}\label{79}
R=\frac{-3a\ddot{a}f^2+2f''f-3\dot{a}^2f-1+f'^2}{a^2f}.
\end{equation}
 By definition
$a>0$ and $\frac{\dot{a}}{a}>0$ \cite{28}, it follows that curves
of $a(t)$ versus $t$ must be concave downward and must reach
$a(t)=0$ at some finite time in the past. Let us recall this time
$t=0$ at which $R=\infty$. In cosmology, extrapolation of the
universe expansion backwards in time yields an infinite density at
finite past. Also, if the strong energy condition \cite{1} is
satisfied, i.e., $\rho+p\geq 0$ and $(\rho+3p)\geq 0$ then $a=0$
at $t=0$ which implies the divergence of scalar curvature
polynomial where $\rho\rightarrow\infty$. This is spacelike
singularity usually called \textit{big bang singularity} or
\textit{initial singularity} \cite{29}.

\section{Summary and Conclusion}

In this paper, we have analyzed the charged perfect fluid collapse
with positive cosmological constant in the Friedmann models. For
this purpose, we have found junction conditions between the
Friedmann models and the Reissner-Nordstr$\ddot{o}$m de-Sitter
spacetime. The junction conditions provide the gate way for the
exact solution of the field equations with interior spacetime
(Friedmann models). The solution of the field equations helps to
discuss the dynamics of the collapsing system as follows:

The acceleration parameter $\ddot{a}/a$, given by Eq.(\ref{14}),
will be zero, positive or negative for
$4\pi(\rho+3p)+E^2-{\Lambda}=0,~4\pi(\rho+3p)+E^2-{\Lambda}<0$ or
$4\pi(\rho+3p) +E^2-{\Lambda}>0$ respectively. The variation of
the scale factor $a(t)$ helps to describe the different stages of
matter in the Friedmann models of the universe \cite{30}. If the
scale factor $a(t)$ is decreasing, i.e., $\dot{a}(t)<0$ there will
be collapsing (contracting) phase. For increasing scale factor
i.e., $\dot{a}(t)>0$ we have the expanding phase while the point
where $\dot{a}(t)=0$ corresponds to bounce point. Consequently,
the Hubble parameter will be $H<0,~H>0$ and $H=0$ for collapsing,
expanding and bouncing phases respectively. Also, we can conclude
the following:
\begin{itemize}
\item The Newtonian force and acceleration of matter
have the same value over the hypersurface $\Sigma$, i.e.,
($-\frac{m}{(af)^2}+(\Lambda+{E^2}-{8\pi}p_c)\frac{(af)}{3})_\Sigma$
(see \cite{23} for detail). In this case, the repulsive force can
only be generated if $\Lambda>({8\pi}p_c-{E^2})$ such that
{${8\pi}p_c>>{E^2}$} over the entire range of the collapsing
sphere. In the case of charged perfect fluid collapse with
Tolman-Bondi spacetime \cite{23} there is no restriction on matter
and electromagnetic field then the results are valid only for
$\Lambda>({8\pi}p_c-{E^2})$ such that ${8\pi}p_c>{E^2}$. It is
clear that in the first case the cosmological constant attains
higher value than the later case. Thus the cosmological constant
plays an effective role to slow down the collapse in the present
case than previous one. In other words, isotropy and homogeneity
of matter causes to introduce resistance against collapse in the
presence of charge.
\item Since the cosmological constant $\Lambda$ is affected by
pressure and electromagnetic field, we can say that
electromagnetic field reduces the effects of $\Lambda$ as compared
to perfect fluid case by putting the restriction on $\Lambda$.
Hence electromagnetic field increases the gravitational collapse
as it decreases the repulsive force produced by $\Lambda$.
\item Two physical horizons (cosmological and black hole horizons)
are found whose area is decreased by cosmological constant and
electromagnetic field. It follows from Eq.(\ref{70}) that both
horizons form earlier than singularity, so singularity is covered
(back hole) and CCH seems to be valid in this case.
\item Time difference between the formation of apparent horizon
and singularity is decreased by electromagnetic field. Thus we can
say that singularity must form earlier than the apparent horizons.
Hence electromagnetic field favors the formation of naked
singularity. But such situation can never occur because
electromagnetic field does not play the dominant role in this
case.
\item It is found that the time difference between the formation of
cosmological (black) horizon and singularity is decreasing
(increasing) function of mass of the collapsing system.
\end{itemize}

\vspace{0.25cm}

{\bf Acknowledgment}

\vspace{0.25cm}

We would like to thank the Higher Education Commission, Islamabad,
Pakistan for its financial support through the {\it Indigenous
Ph.D. 5000 Fellowship Program Batch-IV}.

\end{document}